# Kondo-Resonance Mediated Metal-Insulator Transition in GaAs Embedded with Erbium Arsenide Quantum Dots

W-D. Zhang[*] and E. R. Brown[†]

Departments of Physics and Electrical Engineering

Wright State University,

Dayton, OH, USA 45435

A. D. Feldman, T. E. Harvey, and R. P. Mirin

Applied Physics Division

National Institute of Standards and Technology

Boulder, CO, USA 80305

[*] wzzhang@fastmail.fm

[†] elliott.brown@wright.edu

**Abstract**

We report anomalous critical transport behavior in a GaAs structure containing a dense array of ErAs quantum dots. The structure displays a voltage (electric field)-controlled insulator-to-metal transition and strong hysteresis in the Kondo-like current-vs-temperature characteristic, with critical temperatures as high as 50 K. We attribute this behavior to a strong distributed Kondo resonance between the quantum dots after the Coulomb blockade of the array is lifted. This is consistent with a high sensitivity of the phase transition to a small external magnetic field that we have observed in the Voigt configuration, and a phenomenological model based on the RKKY interaction within a quantum dot and the cooperative Kondo-resonance amongst quantum dots.

**Introduction**

In some metals containing magnetic impurities, such as iron-in-gold, electrical resistivity can display a minimum with respect to decreasing temperature instead of a monotonic decrease as metals typically do.[1] The temperature dependence of the resistivity around the minimum is described with a characteristic $|\log(T/T_K)|$, where $T_K$ is the Kondo temperature. It was Kondo who first showed that this behavior follows from a 3$^{rd}$-order perturbation interaction involving the spin-flip scattering of conduction electrons by localized spins of the magnetic impurities.[2] Abrikosov, Suhl, and Nagaoka showed that the divergent behavior in Kondo's calculation has physical meaning and could be explained by *resonant scattering* at $T < T_K$,[3-5] corresponding to an "anomalous" density-of-states (DoS) near the Fermi level created by the non-perturbative antiferromagnetic spin-exchange coupling. The related Kondo problems —such as the second-order virtual processes and the break-down of perturbation techniques below the Kondo temperature －were further advanced by Anderson's magnetic impurity model, "poor man's" scaling theory, as well as Wilson's renormalization group technique. [6-8]

This Kondo-Abrikosov-Suhl-Nagaoka (abbreviated here as Kondo) resonance phenomenon was also discovered for a single non-magnetic quantum dot about two decades ago. [9-15] However, contrary to the resistive behavior of magnetic atoms in metals, for a quantum dot with an unpaired spin, the electrical conductance increases as temperature drops below $T_K$, varies via the characteristic $|\log(T/T_K)|$ near $T_K$, and saturates at the quantum conductance $G_0=2e^2/h$ (e-electron charge; h-Planck constant) as T approaches zero. Below $T_K$, the magnetic moment of the quantum-dot net spin is screened out by its surrounding conduction electrons, and a spin singlet state is formed. The strong spin exchange coupling between the localized quantum dot electron and the conduction electrons with opposite spin creates a Kondo resonance state *which is always pinned at the Fermi level,* thus generating an enormous DoS. The Kondo resonance provides a favorable conduit for electrons to transit through the binding potential of a quantum dot, which would otherwise act as a scattering barrier or absorbing trap with respect to the electron transport. [9-19] All previous studies on the Kondo effect entailed only one, two or three quantum dots,[20-22] or an ellipse of cobalt atoms, [23] and were observed only at low cryogenic temperatures (<1 K). In this letter, we report a Kondo-resonance behavior in samples containing a large array of rare-earth quantum dots in a GaAs matrix, with $T_K$ in the range ~17-50 K. We discover not only transport properties similar to the one found in a single quantum dot but also a new phase transition phenomenon which involves the cooperation of multiple quantum dots at the macroscopic scale through Kondo resonances.

**Results**

The rare-earth-doped samples were created by molecular beam epitaxy (MBE) of GaAs doped with two different erbium (Er) concentrations: (A) $3.5 \times 10^{17}$ cm$^{-3}$ (control sample, 1.25 µm-thick) and (B) $8 \times 10^{20}$ cm$^{-3}$ (1 µm-thick). These straddle the widely accepted solubility limit of $\sim 7 \times 10^{17}$ cm$^{-3}$ for the incorporation of atomic Er,[24] so Sample B is expected to have many ErAs quantum dots (Fig. 1a) [Methods], and Sample A few or none. The Er and As atoms spontaneously form ErAs quantum dots partly because the NaCl crystal structure of ErAs has nearly the same cubic lattice constant as the zinc blende of GaAs, and hence the quantum dots are energetically stable. Transmission electron microscope (TEM) images of Sample B (Fig.1b) show that the GaAs material maintains high quality without other noticeable defects such as threading dislocations. The diameter of the quantum dots is predominantly distributed in the range of ~2-5 nm (Fig. 1c), the "most likely" diameter being 2.5 nm. The quantum dots are densely packed with a typical center-to-center separation of ~3.4 nm and an estimated concentration of $\sim 5 \times 10^{18}$ cm$^{-3}$. Each quantum dot has a potential-energy profile like that illustrated in Fig. 1d, which is based upon near-infrared photoabsorption data (Supplementary info. I, Fig. S1),[25, 26] and other studies using scanning-tunneling-microscopy.[27] Figure 1a illustrates the structure of the device under test (DUT): two 4.5-mm-wide, 5.0-mm-long ohmic-contact stripes on top of the epitaxial layer separated by a gap of $L_C \approx 0.5$ mm.

To qualify our experimental technique, we started with the DUT Sample A (DUT#1) and measured current (I) vs temperature (T) at different bias voltages ($V_b$), showing typical conduction mechanisms for semiconductors doped with point defects (Figs. 2 a, b).   At 45 V bias, between ~130 K and 300 K, the conductance ($G_A$) displays an Arrhenius behavior with activation energy ~0.13 eV (Figs. S2, S3, and Supplementary info. II).[28]    As the temperature drops from ~130 K to ~95 K, $G_A$ vs. T is fit with Mott variable-range hopping (VRH); [28] below ~95 K but above the "freeze-out" temperature ~53 K, $G_A$ is best fit with Efros-Shklovskii (ES) VRH. [29] Analysis of the 5 V-bias curve yields similar results, but without the ES VRH at low temperatures.

We then characterized DUT Sample B (DUT#2) with the known-high concentration of ErAs quantum dots, ~$5 \times 10^{18}$ cm$^{-3}$.   Figure 3a plots the I-T curve at low bias $V_b$=5 V (similar to 10 V bias data in Figs. S4, S5).    Unlike the piece-wise conduction mechanisms found in Sample A, the entire conductance curve between ~17 K and 300 K is fit better by ES VRH (Fig. 3b) than Mott VRH (Fig. S6), and there is no Arrhenius behavior at the high temperatures.    The ES-type behavior suggests that the DoS for the ErAs quantum dot array may have a soft "Coulomb gap" centered at its Fermi level such that DoS($E_F$) is proportional to $(E - E_F)^2$. [30, 31] As T drops below ~17 K, $I_B$ decreases to the lower limit of the source meter, and the precipitous drop is consistent with the insulating behavior expected from the Coulomb gap.   Both the temperature-decreasing and -increasing curves (Fig. 3a) are practically superimposed over the entire range, contrary to the hysteretic behavior described next.

Fig. 3c shows $I_B$ vs T behavior when the DUT was biased at a higher voltage of $V_b$=45 V while other experimental conditions were kept the same (similar to 50 V bias, Fig. S7). On the temperature-decreasing curve, just below a critical temperature of $T_C$~16.3 K, the current suddenly jumps up almost two orders of magnitude and then tends to a constant value ~$7 \times 10^{-6}$ A, down to at least 4.2 K. Such conductance increase around $T_C$ and then saturation as T approaches zero is qualitatively similar to the Kondo effect that occurs with a single quantum dot.[11-17] However, the increase in Fig. 3c is discontinuous — indicative of a phase transition — whereas the single quantum-dot's logarithmic behavior is *continuous*.

Further characterization of the critical behavior is observed in the temperature - increasing curve of Fig. 3c, which forms a clockwise hysteresis together with the temperature-decreasing curve. The high conductance is maintained approximately constant up to $T_K$~49 K, where a steep decrease occurs. However, the quantitative drop is not as steep as for the temperature-decreasing curve near $T_C$, and is found to have a $|\log(T/T_K)|$ dependence near $T_K$ (Fig. 3d), which is a characteristic feature of the Kondo effect.[11-17]

Next, at a fixed temperature below $T_C$ such as T≈4.3 K, voltage sweeps were performed from low-to-high and then high-to-low, yielding the counter-clockwise hysteresis loop shown in Fig. 4 a. When $V_b$ is less than ~16 V, the device is insulating. As $V_b$ increases, $I_B$ increases monotonically but then displays a sudden increase in slope at $V_M ≈ 35$ V. At a higher critical voltage of $V_{CH} ≈ 51.5$ V, $I_{DUT}$ abruptly jumps up from $4.4 \times 10^{-8}$ A to $8.0 \times 10^{-6}$ A-approximately two orders of magnitude. The abrupt increase, which is consistent with the temperature-dependent curves at fixed biases such as 45V (Fig. 3 c), cannot be cross-gap impact ionization in GaAs, or impact ionization of the quantum dots as the E field in our experiments is too low, <$10^3$ V/cm, compared to typical threshold fields (> $1 \times 10^5$ V/cm). As $V_b$ sweeps down, $I_B$ drops critically at a (lower) voltage of $V_{CL} ≈ 26$ V.

Further, as T is increased from 4.4 K to 5.8 K, $V_{CH}$ ($V_{CL}$) of the current-voltage curves shifts to the right (left) by ~1 V, respectively (Fig. 4 b). $V_{CH}$ rises because more electrons, and thus more current, are required for the establishment of the metallic state to overcome the greater thermal agitation of the higher temperature.

Finally, we measured the I-V behavior at a fixed temperature of 77 K (liquid nitrogen immersion) without and with a magnetic induction of ~1320 Gauss provided by a Nd:Fe permanent magnet mounted in close proximity to a DUT Sample B (DUT#3). The magnetic field was orientated perpendicular to both the current and the DUT substrate (the Voight configuration). As plotted in Fig. 4 c, $V_{CL}$ and $V_{CH}$ move from ≈7.8 and 13.7 V to ≈6.8 and 15.3 V, respectively. Between ~13.7V and ~15.3V, there is a large negative magnetoresistance (>3×10$^4$ ratio), indicating an antiferromagnetic order possibly plays essential role in the electronic transport of the metallic state. This leads us to a magnetic-field dependent qualitative model of the quantum-dot cooperative behavior, as described next.

**Discussion**

One indication of macroscopic cooperative behavior amongst the quantum dots is the electrical conduction in the metallic state of Fig. 3c at 45 V bias. Unlike at 5V (Fig. 3 a), the conduction at 45V is metallic because the conductance is nearly flat approaching the lowest experimental Ts so not described through either Mott or ES VRH among localized states.[32] Extrapolation to 0 K yields an asymptotic current ≈7.0 μA, which in the geometry of the DUT corresponds to a bulk conductivity of 1.7x10$^{-2}$ S/m. We compare this to the minimum metallic conductivity $\sigma_{min}$ determined with the Ioffe-Regel criterion:

$\sigma_{min} \approx 0.03e^2/L$, where L is the mean free path.[28] On first inspection, L would logically be the mean separation between quantum dots. Knowing that the Er fraction is 4.0% by volume, and assuming that the quantum dots all have the most likely diameter of 2.5 nm (Fig. 1 b), we estimate L to be ~3.4 nm, which leads to $\sigma_{min} = 2.2 \times 10^3$ S/m - a value nearly 5 orders higher than the experimental value. However, if we let L be the separation between the contacts $L_C$, we calculate $\sigma_{min} = 1.5 \times 10^{-2}$ S/m - a value lower than the experimental value by only 18%. This supports that the electron conduction in the metallic state is highly cooperative at a macroscopic scale.

Based on the low-temperature data and the magnetic sensitivity, we propose that the cooperative behavior among quantum dots is likely related to collective Kondo resonances. Although only studied experimentally in thin films or nanoparticles, ErAs is known to be antiferromagnetic with a Neel temperature of ≈5 K, and paramagnetic at higher temperatures.[33] The paramagnetism is dominated by the atomic Er which incorporates into the rock-salt structure of ErAs as a trivalent ion $Er^{3+}$ with a configuration of $4f^{11}$,[34] and an effective magnetic moment of ~5.3 Bohr magneton.[33] A profound consequence is that a quantum-dot confined (QDC) electron harbored within an ErAs quantum dot can be polarized through the Ruderman-Kittel-Kasuya-Yosida (RKKY) exchange interaction, because the QDC electron's wavefunction is spread over the entire dot (~2-5 nm) and thus itinerates amongst the much larger number of atomically-localized (~3Å) 4f electrons (Fig. 5 a).[35-37] The spin-polarized QDC electron(s) can be screened by those electrons with *opposite spin* that hopping back and forth to neighboring quantum dots (Fig. 5 b). This is supported by Anderson's scaling theory that antiferromagnetic Kondo coupling is strong.[7] With the formation of the "Kondo cloud", the probability of a "mobile" hopping electron traversing the quantum dot via the Kondo resonance with a "local" QDC electron increases substantially (Fig. 5 c). [11-20] The Kondo resonance happens to not only one quantum dot, but also a dense array of quantum dots, this is because a Kondo resonance state for any single quantum dot is always aligned with the Fermi energy, which must be uniform among quantum dots in thermodynamic equilibrium (Fig. 5 d, e). This "double pinning" effect is key to establish the cooperative Kondo resonances, which produce such a huge amount of $DoS(E_F)$ that a metallic behavior occurs.

Figure 6 illustrates the difference between the $DoS(E_F)$ in the insulating and the metallic states. For the insulating state, the Fermi level falls within a Coulomb gap created by the electron-electron interaction as well as the strong disorder (Fig. 6 a). For the metallic state, the Fermi level is aligned with the sharp peak of the DoS because the correlated Kondo resonances overcome the Coulomb blockades (Fig. 6 b). During the transition, electron spins arrange from random orientation to orderly formation of Kondo singlets. A singlet has total spin S=0, so S is no longer a good quantum number. This means the spin-rotational symmetry of the Kondo Hamiltonian is broken, thus the phase transition is likely second order. [38, 39] To the best of our knowledge, such a Kondo resonance-mediated phase transition hasn't been reported previously in any study of magnetic-impurity-doped quantum-dot systems. [40-45]

The conductance-increasing behavior of our GaAs:Er quantum dot composite bears similarity to that of heavy-fermion compounds such as $U_2Zn_{17}$ and $CeAl_3$.[46-49] And the physical explanation is similar: a global Kondo resonance occurring in a Fermi electron sea. [48-50] A subtle but important difference is that the itinerant electrons in the heavy-fermion compounds are positioned within a conduction band, while for the quantum dot array, the itinerant (i.e. hopping) electrons have energies deep in the bandgap of the host semiconductor. At lower bias, too few electrons are available so that either the local magnetic moments of quantum dots are not screened enough or the quantum dots are too de-occupied. Thus ES-type hopping among quantum dots is the primary conduction mechanism. This explains the abrupt voltage-controlled insulator-to-metal transition that we have observed.

In conclusion, our results show that a semiconductor (GaAs) matrix embedded with rare-earth-bearing quantum dots (ErAs) can display an insulator-to-metal transition that is a strong function of bias voltage, temperature, and magnetic field. Our qualitative model proposes that with these quantities increasing or decreasing a critical transition from a Mott insulator to a metallic state occurs via a macroscopic Kondo-resonance. Correspondingly, the density-of-states at the Fermi level can be switched from a "null" Coulomb gap to a "sharp" Kondo resonance peak. Hence, we demonstrate that this type of magnetic-quantum dot-in-semiconductor composite has very unusual electronic, thermodynamic and magnetic properties.

**Methods**

**Material Growth** The growth was carried out by molecular beam epitaxy on 76.2 mm diameter semi-insulating GaAs substrates at a temperature of $\approx 600°C$ as measured using band edge thermometry. The GaAs growth rate was about 0.65 monolayers per second. The Er flux was calibrated previously using secondary ion mass spectroscopy on a separate sample.

**Device Structure** Two-terminal devices were fabricated on the epilayer of rare-earth arsenide quantum-dot composite. The electrodes were ~200 nm thick AuGe deposited with thermal evaporation and then annealed in a nitrogen environment at 465°C for 30s.

**Test Method** Device under test (DUT) was mounted on the cold finger of a Gifford-McMahon closed-cycle-He refrigerator. Apiezon grease was applied between the DUT and the cold finger to ensure excellent thermal conduction. Temperature was measured with a calibrated silicon diode, and then recorded with a temperature controller. All currents were measured with a voltage-current source meter, which has a minimum range limit 10 pA. The compliance of the source meter was set 4 mA for the safety of DUTs. A Matlab script was programmed to collect readings from the temperature controller and the source meter automatically. All the measurements were conducted in dark environment.

**Acknowledgments:**

We thank for a research grant from the U.S. Army Research Laboratory and the U.S. Army Research Office managed by Dr. Joe Qiu under contract number W911NF-12-1-0496.

Official contribution of the National Institute of Standards and Technology; not subject to copyright in the United States.

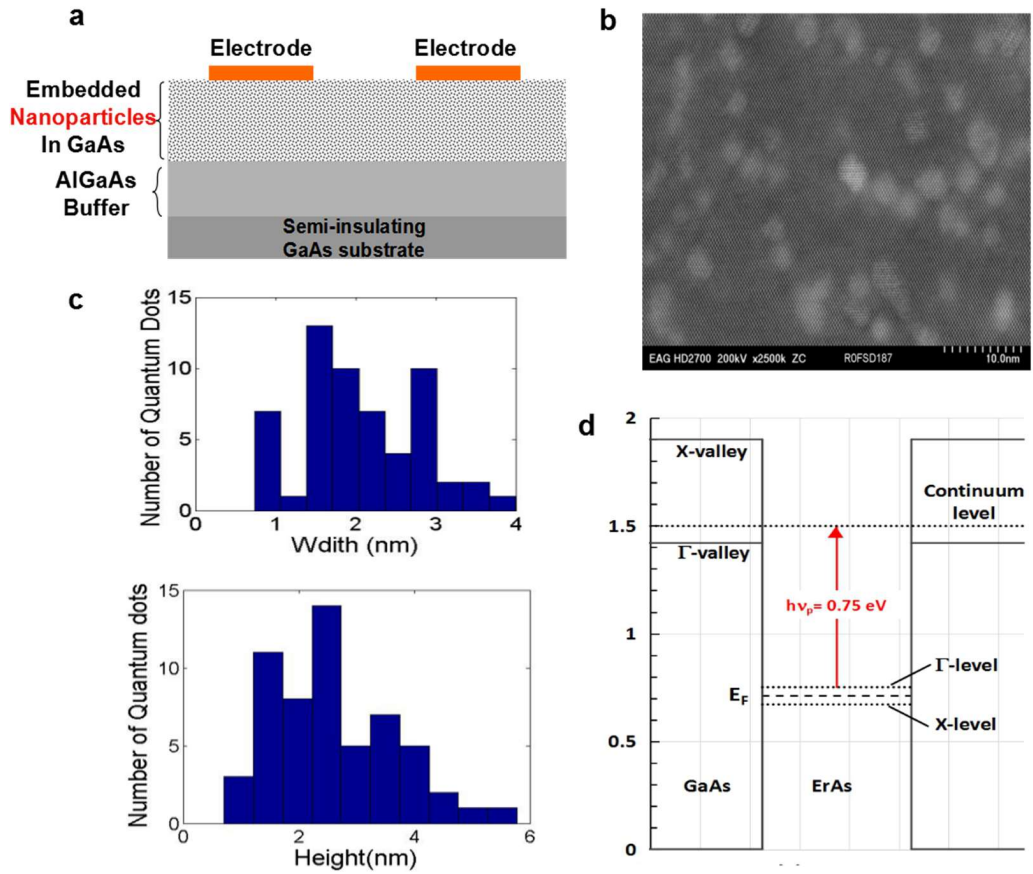

**Figure 1** **a**, An illustration of the GaAs:Er epilayer and the device structure under test. **b**, The TEM imaging-side view-of the GaAs:Er eiplayer confirms the mixing of quantum dots into GaAs. Those light-tone disks indicating the presence of ErAs quantum dots are consistent with Er being much heavier than Ga. **c**, Histograms on the size distribution of quantum dots were generated using an object-recognition software (icy, http://icy.bioimageanalysis.org). **d**, the potential well of quantum dot was constructed with the 1550 nm optical absorption, photoconductivity in reference 25, and other results from references 26, 27. The Fermi level for the ErAs quantum dot array is located near the middle of the bandgap of GaAs (~0.7eV).

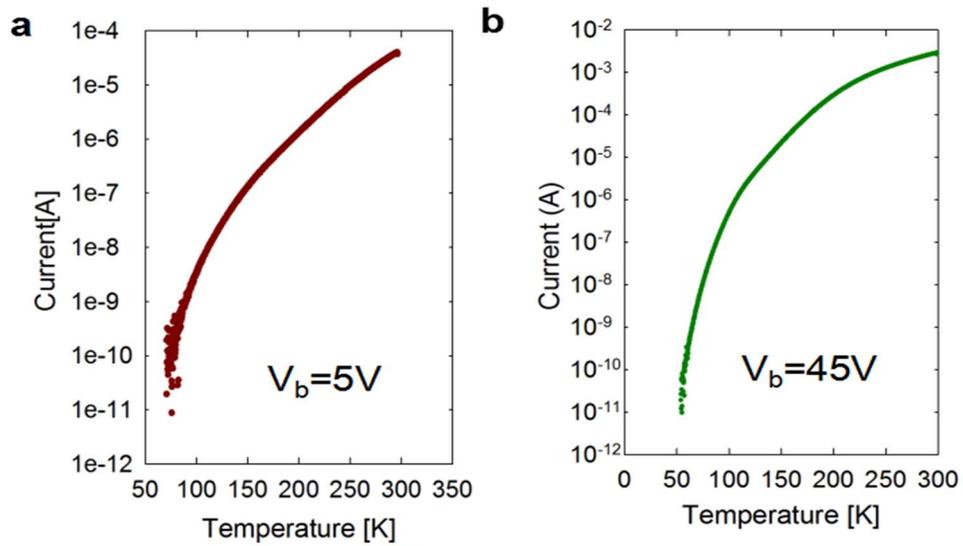

**Figure 2** The current vs temperature in DUT Sample A (control sample). **a,** at a fixed bias of 5 V; **b,** at 45 V bias.

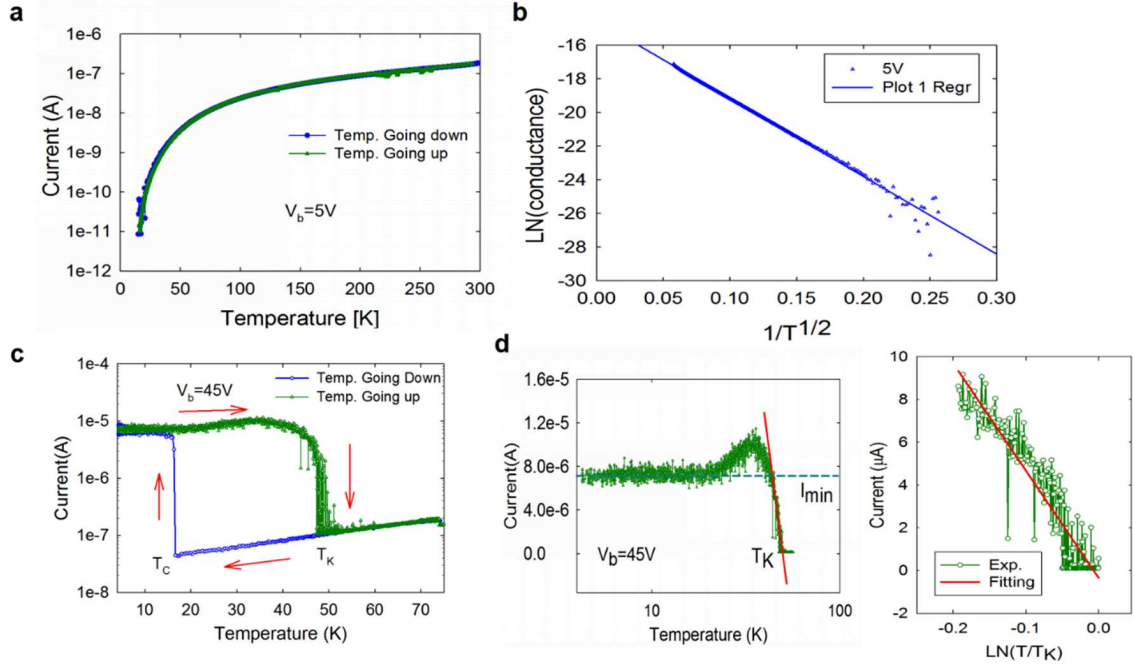

**Figure 3** I-T behavior for DUT#2 (Sample B). **a,** The current vs. temperature curve at a bias of 5V. **b,** The fitting of the experimental data in **a** with the Efros-Shklovskii variable-range hopping. The fitting equation is log(conductance)=-99515-29.6 $1/T^{1/2}$, and $R^2$=0.9849. **c,** The current vs. temperature when the DUT was biased at 45V. On the temperature-decreasing curve at $T_C$~16.3 K, the current jumps up almost two orders of magnitude, and saturates as the temperature decreases toward zero. On the temperature-increasing curve, the high conduction is maintained well above $T_C$, yielding a strong hysteresis in the I vs T curve. The upper edge of the hysteresis region occurs at $T_K$ ~ 49 K. **d,** By isolating the temperature-increasing curve of **c**, the fall-off of the conductance is fit with $|\log(T/T_K)|$ (red line), which is a characteristic of the Kondo effect, as T approaches $T_K$. The fitting of current (μA) vs. $|\log(T/T_K)|$ in the range of ~40.0K-49.7K is plotted on the right side of **d**. The fitting equation is y=-0.34-50.18×log(T/$T_K$), and $R^2$=0.8946.

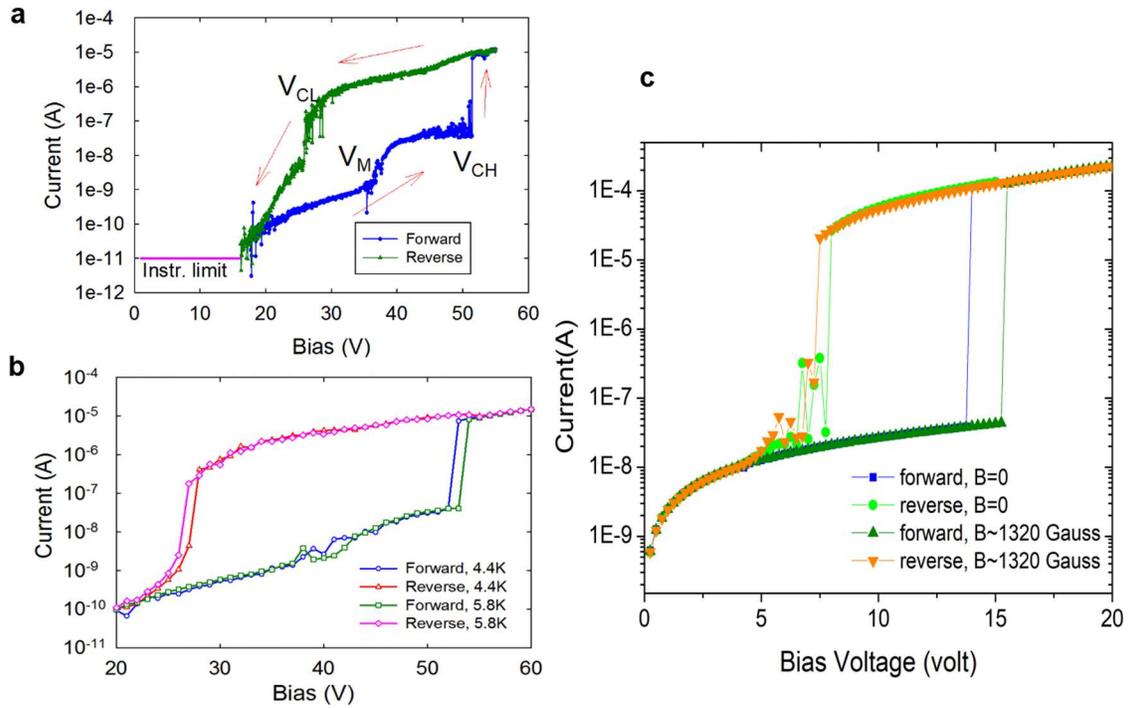

**Figure 4**  **a** Current vs bias voltage (of DUT#2) sweeps vs both forward and reverse at T=4.3 K with critical voltages at $V_{CH}$ and $V_{CL}$, respectively.  The curves display a pronounced hysteresis loop – a telltale sign of phase transitions. **b**, Current vs bias voltage (of DUT#2) sweeps at T=4.4 K and 5.8 K, respectively. $V_{CH}$ shifts to the right by ~1 V while $V_{CL}$ shifts to the left by ~1V as T changes from 4.4 K to 5.8 K.  **c** Current vs bias voltage (of DUT#3) at T=77 K showing that the effect of magnetic induction of ~1320 G is to shift both $V_{CL}$ and $V_{CH}$ in voltage.

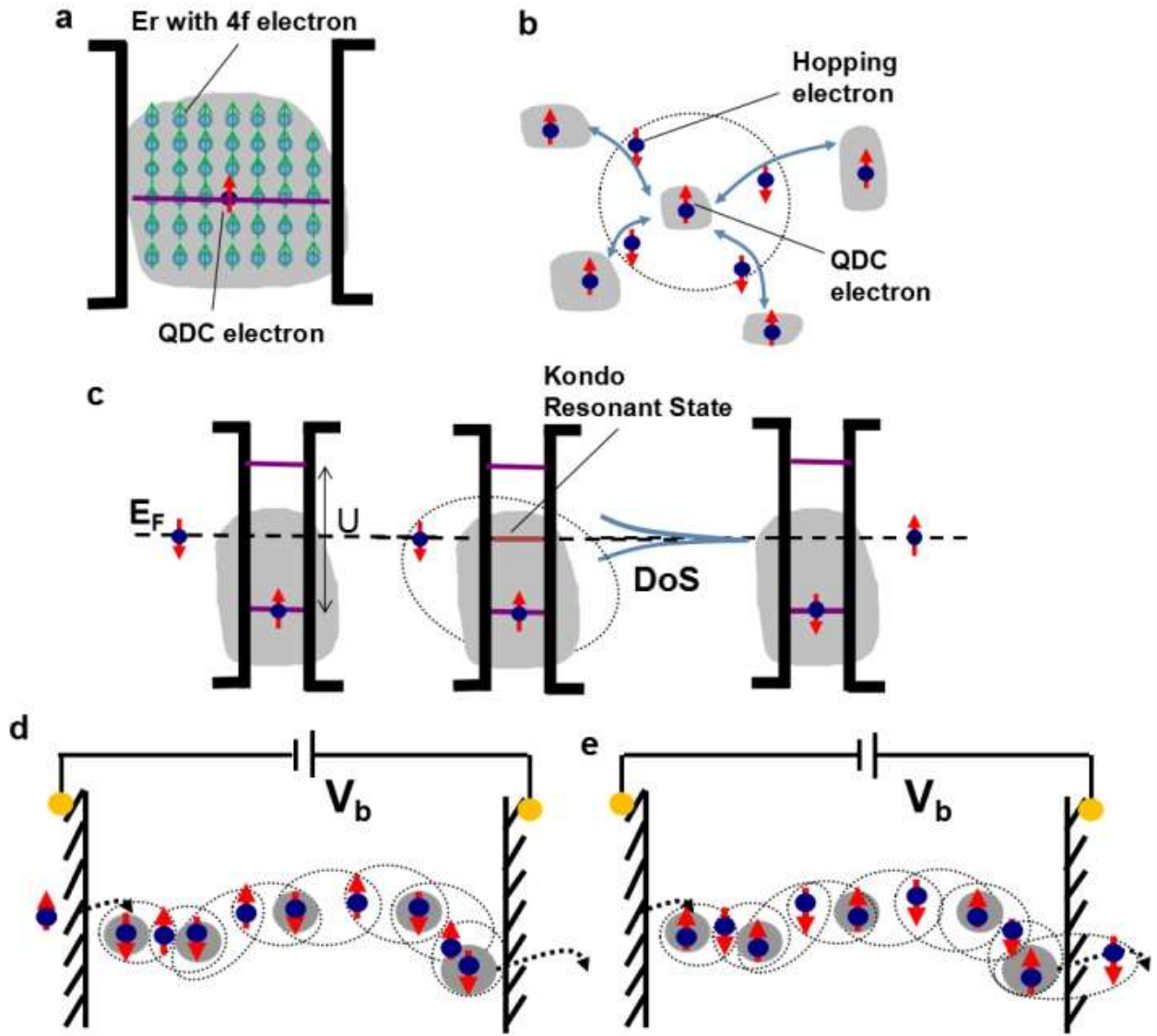

**Figure 5** Kondo resonance mediated insulator-to-metal transition. **a**, A possible scenario that an QDC electron is polarized through the RKKY interaction with the atomically-localized Er ions within the quantum dot. **b**, The spin-polarized QDC electron of an ErAs quantum dot is quenched by hopping electrons in its neighborhood. A Kondo spin singlet is formed between the polarized QDC electron and a hopping electron. **c**, Kondo resonant tunneling occurs due to the strong spin-flip exchange interaction between the QDC electron and one of the hopping electrons. The Kondo resonant state is always aligned with the Fermi level. Since the Fermi level is a global scale, Kondo resonance can take place collectively to many quantum dots. Thus, an enormous density-of-states condenses at the Fermi level, leading to the insulator-to-metal transition. **d** and **e**, A possible trajectory illustrates how electron conducts from one terminal to the other through the domino of Kondo resonance tunneling.

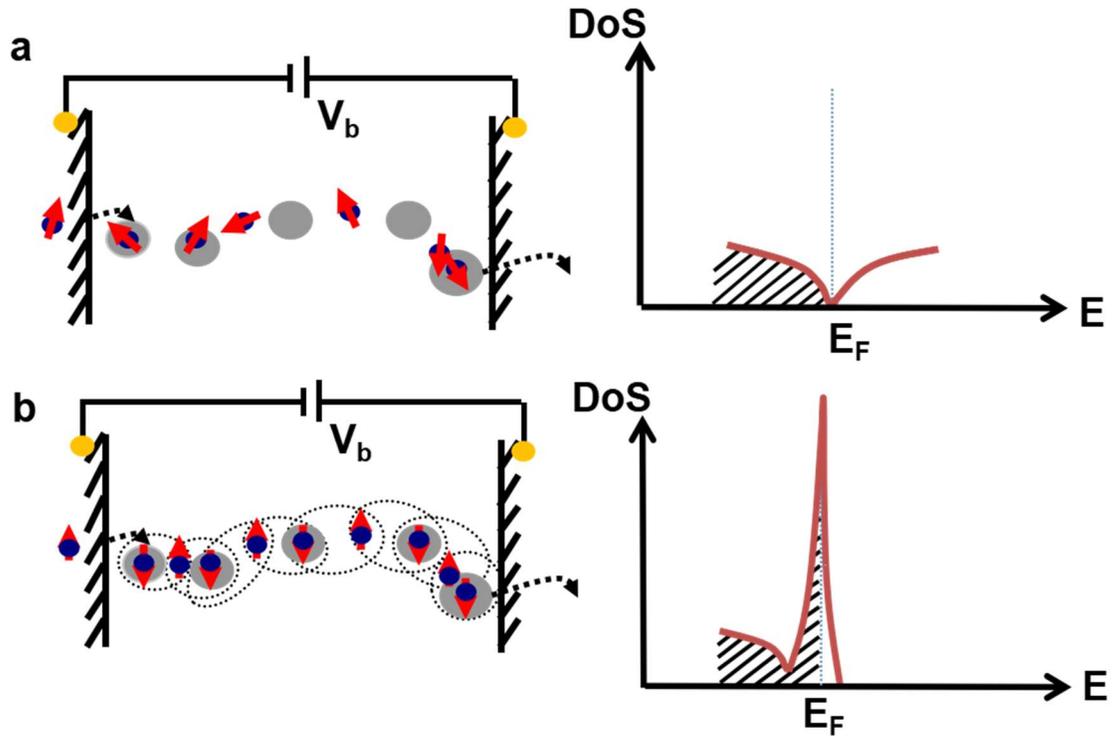

**Figure 6 a**, An illustration on the DoS of the insulating state under a low bias voltage. The Fermi level is located within the Coulomb gap due to the strong electron-electron interaction. **b**, An illustration on the DoS of the metallic state under a high bias voltage. The Fermi level is aligned with the center of a sharp DoS due to the correlated Kondo resonances, which lift off the Coulomb blockades.